\documentclass[pre]{revtex4}% Physical Review B

\usepackage{graphicx}% Include figure files
\usepackage{bm}% bold math

\begin{document}
\title{{\sc USHER}: an algorithm for particle insertion in dense fluids}
\author{R. ~Delgado-Buscalioni}
\email[]{R.Delgado-Buscalioni@ucl.ac.uk}
\affiliation{Centre for Computational Science, Dept. Chemistry, University College London, 20 Gordon Street, London WC, U.K.}
\author{P. V. Coveney}
\email[]{P.V.Coveney@ucl.ac.uk}
\affiliation{Centre for Computational Science, Dept. Chemistry, University College London, 20 Gordon Street, London WC, U.K.}

\date{\today}

\begin{abstract}
The insertion of solvent particles in molecular dynamics
simulations of complex
fluids is required in many situations involving open systems, but this
challenging task has been scarcely explored in the literature.  We
propose a simple and fast algorithm ({\sc usher}) that inserts the new
solvent particles at locations where the potential energy has the
desired prespecified value.  For instance, this value may be set equal
to the system's excess energy per particle, in such way that the
inserted particles are energetically indistinguishable from the other
particles present. During the search for the insertion site, the {\sc usher}
algorithm uses a steepest descent iterator with a displacement whose
magnitude is adapted to the local features of the energy landscape.  The
only adjustable parameter in the algorithm is the maximum displacement and we
show that its optimal value can be extracted from an analysis of the
structure of the potential energy landscape.  We present insertion
tests in periodic and non-periodic systems filled with a Lennard-Jones
fluid whose density ranges from moderate values to high values.
\end{abstract}


\maketitle

\section{\label{int}
Introduction}

Many dynamical processes of chemical, biological and physical interest
occur in open systems where matter and energy are exchanged with the
surroundings. The main focus of attention has been in (Grand Canonical)
Monte Carlo algorithms, which are particularly suited for the study of
equilibrium states. However more recently there has been significant
attention focused on molecular dynamics (MD) algorithms adapted for
open systems. One of the biggest challenges faced in the investigation
of such phenomena by MD simulation is the problem of efficient
insertion of solvent particles in dense liquids. Indeed, as the scope
and scale of MD increases, a growing variety of applications and
methods are in need of {\em on the fly} particle insertion algorithms
(see Refs. \cite{Pet92,Pet_more,waterins,Flek00,hybrid} and references
therein). Among those, a particularly relevant family of methods are
hybrid schemes that couple the particle domain to an outer region
described by continuum fluid dynamics \cite{Flek00,hybrid}. The
present method was devised for use in such hybrid schemes, but we
believe its application within molecular simulation may prove to be
more widespread.

%MODIFICATION
The  problem of  inserting  a solvent  molecule  in a  dense fluid  is
commonly  encountered  in Grand  Canonical  Monte  Carlo methods,  for
instance  in  Gibbs  ensemble  calculations for  phase  equilibria  or
evaluation  of the chemical  potential.  A  number of  techniques have
been      proposed     to      overcome     this      problem     (see
\cite{ExtendGibbs,Cav-Gibbs}  and references therein).   For instance,
cavity-biased procedures  search for domains  within the fluid  with a
small  local value  of  number  density, as  these  cavities are  more
susceptible  to accommodate a  new molecule.  In doing  so, a  bias is
introduced and, according to the  rules of MC simulation, the bias has
to be precalculated and corrected so that the scheme adheres to detail
balance. Even so, recent comparisons  showed that if the molecules are
smaller than the  mean size of the cavities, GCMC  is nearly ten times
faster than GCMD \cite{Mez99}. We believe that the insertion algorithm
proposed  here may  be  used to  improve  the efficiency  of the  GCMD
schemes.

The acme of a particle insertion protocol for MD is one that, in just
a few iterations, is able to place the new particle within the
required subdomain of the simulation space at a site where the
potential energy takes exactly the desired value. This last condition
ensures that no extra energy is introduced into the system, and
therefore such an insertion algorithm would not require
thermostatting after each insertion.  Indeed, even for moderate liquid
densities, these are difficult requirements and the few insertion
protocols proposed in the literature \cite{Pet92,waterins} are far
from fulfilling them.

For instance, Goodfellow {\em et al.} \cite{waterins} introduce
solvent (water) molecules in the cavities of proteins to investigate
their structural stability. Once the protein's cavities are found, the
insertion protocol consists of several steps that involve operations
over the whole system. Solvent molecules are introduced with arbitrary
orientation and locations within the selected cavity. As a
consequence, the energy of the system increases sharply after each
solvent molecule insertion and, to allow its relaxation, two hundred energy
minimization steps of the whole system (protein + water) are then
performed, followed by a one picosecond molecular dynamics simulation.
This expensive insertion procedure, which involves substantial
alterations of the microscopic dynamics, could be avoided if the
solvent particle were initially introduced at the desired
potential energy site.

The work by Pettitt and coworkers, towards the Grand Canonical
Molecular Dynamics method \cite{Pet92} (see also \cite{Pet_more} for
further developent and applications), is an example of an open system
MD simulation which does takes care of the potential energy at the
insertion site. In their method, new particles contribute a fractional
number to the total number of particles. These fractional or scaled
particles must be inserted at positions where the potential energy is
equal to that of the former (added or deleted) fractional particle. As
explained in Ref. \cite{Pet92}, the authors first use a grid method to
slice the MD domain into a number of boxes that is the same as or a
little larger than the total number of particles. The most favourable
boxes (with the least number of neighbours) are selected as candidates
to add the new particle. Then the new solvent candidates are placed
within each of these boxes and two hundred possible molecule
orientations of these new solvent candidates are computed.  For each
box, the orientation that yields the potential energy closest to the
desired value is chosen and a first steepest descent procedure of
O(10) steps follows.  If this does not lead to any site with the
desired insertion energy, they finally perform a large steepest
descent procedure with at least 100 steps on the most favourable
box. As the authors mention, this numerically expensive 
protocol still yields numerical errors that can disturb the
system \cite{Pet92}.

Our main concern is the insertion of solvent particles in the
framework of a hybrid (particle-continuum) scheme.  In recent work we
proposed a hybrid sheme that is able to deal not only with momentum
but also with mass and energy exchange between the continuum (C) and
the particle regions (P) \cite{hybrid}.  In particular, particles need
to be inserted in the overlapping $C\rightarrow P$ regions where the
C-fluxes are imposed on the P-domain.  In a real liquid (with
interacting potential energy), mass and energy exchanges are strongly
coupled and we showed that, in order to balance the energy flux, the
new particles have to be inserted at positions where the potential
energy equals the value prescribed by the continuum domain.  In that
work \cite{hybrid} we used a particle insertion algorithm (called {\sc
usher}) which is able to tackle this task in a rather efficient way
(see \cite{hybrid} for a brief description).  Further research has led
to an enhanced version of the {\sc usher} algorithm. Here we shall
describe this new version of the {\sc usher} protocol and, for the
sake of consistency, we shall briefly review the one one presented in
Ref. \cite{hybrid}.

The rest of the article proceeds as follows.  We first formulate the
root-finding problem in Sec. \ref{S2}.  In Sec.\ref{des} we describe a
reference scheme against which we compare the {\sc usher} algorithm
and then the {\sc usher} protocol. Insertion tests are described in
Sec. \ref{tests} and the results are discussed in Sec. \ref{results}.
In Sec. \ref{discusion} we present an analysis of the potential energy
landscape that proves to be very useful for the optimisation of the
algorithm's parameters. Finally, conclusions and directions for future
research on insertion algorithms are given in Sec. \ref{con}.

\section{\label{S2} The insertion problem}

We consider a set of $N$ particles inside a box of volume $V$ which
interacts via pair potentials, $\mathcal{V}(r)$.  At any instant the
potential energy can be defined at any point $\mathbf{r}$ by evaluating
$U(\mathbf{r})=\frac 12 \sum_{i=0}^{N} 
\mathcal{V}(|\mathbf{r}-\mathbf{r}_i|)$.  The force that a test
particle would feel at any point can be measured by
$\mathbf{f}(\mathbf{r})= -\nabla U(\mathbf{r})$.  In this work, we
consider a Lennard-Jones fluid whose interparticle potential
$\mathcal{V}(r)=4(r^{-12}-r^{-6})$ is written in the usual units of
length (the effective radius $\sigma$) and energy (the potential well
$\epsilon$).

The objective of the algorithms presented below is to find a position
$\mathbf{r}_0$ for which the potential energy equals a prescribed
value, $U_0$; therefore $U(\mathbf{r}_0)=U_0$. In most practical
situations a less stringent requirement needs to be fulfilled, namely
$\left<U(\mathbf{r}_0)\right>=U_0$, where brackets denote an average
over a certain (small) number of insertions. 

Even for a simple system such as the Lennard-Jones fluid, the
structure of the energy landscape is very complex, with large energy
gradients, and complicated energy isosurface shapes. A typical energy
distribution along the whole space spreads over several order of
magnitudes, but for the typical (moderate) temperatures usually
considered in applications, the particles need to be placed at
positions with extremely low energies compared with the range of the
energy distribution.  The result is that the mean specific excess
energy resulting from the equation of state
$u_{eos}=U_{eos}(\rho,T)/N$ is a very low energy compared with the
typical energies found at any arbitrary point of the space.  As the
fluid particle density increases the situation worsens, reflecting the
fact that particles tend to reside within deep potential wells.  The
relation of the chemical potential $\mu$ with density $\rho$ shows
clear evidence of this fact. At moderate densities the value of $\mu$
is close to $u_{eos}$ but, above a certain density, $\mu$ steeply
increases above $u_{eos}$, meaning that the typical energy needed to
insert a particle becomes much larger than the mean potential energy
per particle.

Therefore if one needs to insert particles at positions with energies
close to the mean excess energy per particle, $u_{eos}$, one needs to
find extremely low energy sites, particularly in dense systems.  The
main problem to be faced is that the energy landscape presents many
energy `` holes'' whose local minima range from intermediate to
large-energy values. Here, we define a ``hole'' as a region of space
enclosed by an isosurface of energy in such way that $\nabla
U(\mathbf{r} )\cdot \mathbf{n}>0$ at the hole surface, where
$\mathbf{n}$ is the (outward) normal surface vector.  Usually, these
holes act as traps for the widely used energy-minimisation algorithms
based on the standard steepest descent or conjugate gradient methods
\cite{MDbook}. As a matter of fact, we soon discovered that it was
very inefficient to move downhill over the potential energy by means
of any of the standard versions of the steepest descent method used in
molecular simulations (see for instance, Ref.\cite{MDbook}).
%For instance, when using line minimization to find the lowest energy
%site along the steepest direction, but, due to the extremely uneven
%surfaces we have to deal with, by using a standard method it becomes
%very difficult to find the correct displacement ensuring the
%bracketing of the next minima along the steepest direction.
The purpose of the present study is to present a steepest-descent-like
iterative procedure that can sort out the intermediate-energy
holes. To this end the {\sc usher} algorithm does not rely on line
minimisation along the steepest descent direction \cite{numrec},
but instead on a displacement size which is adapted
on the fly, according to the local topology of the potential
energy landscape.  Another advantage is the facile implementation of
the {\sc usher} code.

\section{\label{des} Description of the algorithms}

We shall now describe some general aspects of the problem of particle
insertion and the common features of the algorithms concerned.

In any insertion procedure the first step is to place the new particle
at a starting position $\mathbf{r}^{(0)}$. In all the tests presented
in Sec. \ref{tests}, we chose $\mathbf{r}^{(0)}$ at random.  We also
tried to select $\mathbf{r}^{(0)}$ according to a cavity-biased
procedure (as in Ref. \cite{Pet92}). 
As explained in Sec. \ref{int}, this procedure incurs a
number of operations of $O(N)$ prior to the insertion algorithm itself
and we found that, when using the {\sc usher} algorithm presented
below, it did not reduce the total number of iterations
with respect to the (much cheaper) random choice.

During successive iterations the iterator's position is moved
according to the update rule which, in general, may be a function of
the mechanical quantities at the previous iteration,
$\mathbf{r}^{(n+1)}= \mathbf{r^{(n+1)} }
\left(\mathbf{r}^{(n)}, U^{(n)}, \mathbf{f}^{(n)}\right)$.  
The search terminates if the new position
$\mathbf{r}^{(n+1)}$ is a site with the desired energy. This is
determined by the following condition
\begin{equation}
\label{errmax}
|\xi|^{(n+1)} <\xi_{\max},\;\;\;\mathrm{with}\; \xi^{(n+1)}\equiv \frac{U^{(n+1)}-U_0}{|U_0|}
\end{equation}
where $\xi_{\max}$ is a pre-determined parameter, namely the
half-width of the interval of the accepted energies around $U_0$, and
$\xi$ is defined as the relative difference of the potential
energy $U^{(n+1)}$ at the $(n+1)$ iteration with respect to the desired
value $U_0$.

Finally, once the new particle is correctly inserted, the force that
it exerts on its neighbours is calculated and its velocity is also
assigned. This velocity is drawn from a Maxwellian distribution with
the desired temperature $T$ and the desired mean velocity,
$\left<\mathbf{v}\right>$
\begin{equation}
P(\mathbf{v})=\left(\frac{1}{2 \pi
mkT}\right)^{3/2}\exp\left(\frac{-m(\mathbf{v}
-\left<\mathbf{v}\right>)^2}{2mkT}\right).
\end{equation}

While the algorithm is guiding a new particle to a correct location,
the positions of all the other particles remain frozen.  This means
that one insertion iteration only involves the evaluation of the force
on a single particle (that is, the force exerted by all the particles
at the site $\mathbf{r}^{(n)}$).

The starting position determines whether the following iterations will
have to be downhill (if $U^{(0)}>U_0$) or uphill (if
$U^{(0)}<U_0$).  A simple way to unify both cases in a single scheme
is to rescale the potential energy as $U(\mathbf{r}) \rightarrow
sgn\,U(\mathbf{r})$, where $sgn\equiv
\frac{U^{(0)}-U_0}{|U^{(0)}-U_0|}$.  By doing so the forces
$\mathbf{f}=-\nabla U$ are also redefined and, in particular, a case
with $sgn=-1$ then implies that the redefined forces point uphill of
the (unscaled) potential energy throughout the entire course of that
particular particle insertion. In the following presentation we shall
assume that the energy and force field are already rescaled as
$sgn\,U(\mathbf{r})$ and $sgn\,\mathbf{f}$, so we will not explicitly
include $sgn$ in the equations.

Durin the iterative process the algorithm will encounter three
different situations which may require a separate treatment (for
instance, different update rules).  We denote these situations as follows
as: {\em downhill} move, $U^{(n+1)}< U^{(n)}$; {\em uphill} move, $U^{(n+1)}>
U^{(n)}$ and {\em confinement}, $U^{(n+1)}< U_0 < U^{(n)}$.  
In an optimal insertion one expects to keep going downhill
(with respect to the rescaled potential energy) until the confinement
is attained. Then, the desired location lies within the segment
$\delta \mathbf{r}^{(n)}=\mathbf{r}^{(n+1)}-\mathbf{r}^{(n)}$ and 
can be determined by means of standard one-dimensional (1D)
root-finding algorithms (such as the Newton-Raphson or bisection
methods). 

The most problematic iteration corresponds to the uphill move 
and it deserves some discussion.  To illustrate the $U^{(n+1)}>
U^{(n)}$ scenario we refer to the energy landscape shown in
Fig. 1. Even for moderate densities (Fig. 1 corresponds to
$\rho=0.6$), the low energy regions conform to a complex tube-like
structure.  The insertion algorithm will have to usher the new
particle into these energy tubes before arriving at a correct
location.  An uphill iteration
may arise when the iterator faces either of
two features of the energy landscape: 
intermediate-energy holes or sharp bends (including saddle points).  
Note that both kind of features induce
completely different decisions.  The best thing to do when
encountering an energy trap is to give up the search and restart from
another initial position $\mathbf{r}^{(0)}$.  By contrast, if a bend 
in the
energy landscape leads to a low energy valley, it may be worthwhile to use
an update rule that can efficiently deflect the iterator's trajectory.
Unfortunately, once an uphill move occurs it is not possible to
distinguish between these two features within only one iteration. 
On the other hand, the number of uphill iterations rapidly increases
as the displacement $|\mathbf{r}^{(n+1)}-\mathbf{r}^{(n)}|$
is made larger than a specified maximum threshold.
In fact, an important issue for the algorithm design is first
to estimate this threshold and then, to determine
the best decision to take upon an 
uphill move (see Sec.  \ref{discusion}).

It is also convenient to introduce a restart condition in order to
avoid any possible stagnation of the algorithm around energy holes.
In particular, if $n>n_{\max}$ the search is restarted from another
initial position $\mathbf{r}^{(0)}$.  Particularly at high densities
(typically above $0.75)$, the overall number of iterations is 
sensitive to $n_{\max}$.  A very large value of $n_{\max}$ corresponds
to many unsuccessful and time-consuming iterations, while a value for
$n_{\max}$ that is too small prevents most of the potentially successful
trials from terminating successfully.  We found that the best
compromise between these two extremes is to make $n_{\max}\sim 0.8
\left<n\right>$, where $\left<n \right>$ is the number of iterations
averaged over a certain number $(\sim 20)$ of insertions.  The value
of $n_{\max}$ depends on the density.  It may be determined from an
initial test-run, or alternatively reassigned on the fly
according to the value of $\left<n\right>$ determined during the simulation.

In the remainder of this section we first define a ``reference'' scheme 
against which we can then discuss and compare the {\sc usher} algorithm
in more detail.

\subsection{The reference scheme
\label{ref}}

In order to better understand the behaviour of the {\sc usher}
algorithm it is helpful to compare its performance with a reference
scheme based on a combination of well established methods widely used
in the literature for root-finding and energy minimisation.  
While moving downhill, the
reference scheme uses a basic steepest descent step with a fixed
displacement $\Delta s_{1}$. The update rule being is
\begin{equation}
\label{rs1}
 \mathbf{r}^{(n+1)}=\mathbf{r}^{(n)} + \frac{\mathbf{f}^{(n)}}{f^{(n)}}\Delta s_{1},
\end{equation}
where, according to standard notation,
$f^{(n)}$ is the modulus of $\mathbf{f}^{(n)}$.

If an uphill move is made, the reference scheme will first try to
deflect the iterator's trajectory in order to adapt itself to a possible bend
in the potential energy surface. By construction of the update rule,
Eq. (\ref{rs1}), the potential energy decreases locally at
$\mathbf{r}^{(n)}$ in the direction $\delta \mathbf{r}^{(n)}\equiv
\mathbf{r}^{(n+1)}-\mathbf{r}^{(n)}$.  Therefore if
$U^{(n+1)}>U^{(n)}$ there must exists a location $\mathbf{r}_m$ where,
$U(\mathbf{r}_m)= \min\left\{U(\mathbf{r}_{\lambda})|
\mathbf{r}_{\lambda}=\mathbf{r}^{(n)}+ \lambda \delta \mathbf{r}^{(n)}
\right\}$.  The reference scheme finds the position $\mathbf{r}_m$ by
means of a line-minimisation of the potential energy along the
segment $\delta \mathbf{r}^{(n)}$ (see \cite{numrec} for details). 
The new position is then recalculated by a steepest descent step
starting from $\mathbf{r}_m$ and with a displacement $\Delta s_2$;
\begin{equation}
\label{linmin}
\mathbf{r}^{(n+1)}=\mathbf{r}_m + \frac{\mathbf{f}_m}{f_m}\Delta s_2.
\end{equation}
The line-minimisation itself requires an inner iterative procedure
(see \cite{numrec}). In view of the narrowness of the
potential-energy tubes, we used no more than three iterations for the
estimation of $\mathbf{r}_m$. Better estimates of $\mathbf{r}_m$ do
not improve the efficiency, but instead lead to a larger number of
force evaluations in the overall scheme.  When a local minimum of
potential energy is found, the iterator's position will bounce back,
moving subsequently upwards and downwards in energy as it hits the
walls of the energy hole.  To avoid this situation, after several
(typically three) consecutive uphill iterations the scheme determines that
it has been trapped in a local energy minimum and consequently
restarts the search from another initial position $\mathbf{r}^{(0)}$.
Finally, once the root has been confined in the segment $\delta
\mathbf{r}^{(n)}=\mathbf{r}^{(n+1)}-\mathbf{r}^{(n)}$, it is expressed
as $\mathbf{r}_0=\mathbf{r}^{(n+1)}+\lambda_0 \delta
\mathbf{r}^{(n)}$, with $\lambda_0$ a real number in
$(0,1)$.  To find $\lambda_0$, the reference scheme uses a 1D
root-finding algorithm which combines the Newton-Raphson method with
the robust bisection method to ensure confinement in case of a failure
of the Newton-Raphson step (due, for instance, to $f^{(n)}\sim
0$). It typically took less than three iterations to calculate
the value of $\mathbf{r}_0$.  The optimum choice for the parameters
$\Delta s_1$ and $\Delta s_2$ is presented in Sec. \ref{discusion}.

\subsection{\label{usher}  The {\sc usher} scheme}

The basic idea of the {\sc usher} insertion algorithm is to use an update
rule to move downhill that can adapt the iterator's displacement 
according to the local topology of the low energy landscape.
This is reflected in the following update rule:
\begin{equation}
\label{ush}
\mathbf{r}^{(n+1)}=\mathbf{r}^{(n)} + \frac{\mathbf{f}^{(n)}}{f^{(n)}} \delta s^{(n)}.
\end{equation}
Equation (\ref{ush}) is essentially a steepest descent scheme with
an displacement $\delta s^{(n)}$ that depends on the iterator's
position. The success of the method resides in a judicious choice of
$\delta s^{(n)}$.  Optimal performance was obtained using the
following expression for $\delta s^{(n)}$, which depends on both the
local potential energy $U^{(n)}$ and force $\mathbf{f}^{(n)}$,
\begin{equation}
\label{ds}
\delta s^{(n)} = \left\{
\begin{array}{lr}
\Delta s_{ovlp},  & \mathrm{if }  \; U^{(n)}>U_{ovlp} \\
\min\left(\Delta s ,\displaystyle{\frac{U^{(n)}-U_0}{f^{(n)}}}\right),& \mathrm{if}  \; U^{(n)}<U_{ovlp}.
\end{array}
\right.
\end{equation}

The best way to illustrate how the adaptive displacement of
Eq. (\ref{ds}) works is to describe how the {\sc usher} scheme
performs one insertion.  As long as the starting position
$\mathbf{r}^{(0)}$ is chosen at random, there is a large chance of
overlap with a pre-existing particle, leading to a very large value of
$U^{(0)}$.  The displacement $\Delta s_{ovlp}$ quoted in the first
line of Eq.(\ref{ds}) can be constructed to remove the overlap in
(typically) one iteration.  For this reason, $U_{ovlp}$ is chosen to be
a very large energy representing an overlap position, say $U_{ovlp}
\sim 10^4$.  As the hard-core part of the interparticle potential goes
like $4r^{-12}$, the distance from a site with energy
$U^{(n)}>U_{ovlp}$ to the centre of the overlapped particle is
$r=(4/U^{(n)})^{1/12}$.  Therefore by choosing $\Delta
s_{ovlp}=r_{\sigma} - (4/U^{(n)})^{1/12}$, we can guarantee that the
next iterator's position $\mathbf{r}^{(n+1)}$ will be moved a distance
$r_{\sigma}$ away from the centre of the overlapped particle and, by
virtue of Eq. (\ref{ush}), in a direction of lower potential energy.
The value of $r_{\sigma}$ should be close to or slightly smaller than a
characteristic contact distance between particles (e.g. the
distance given by the maximum in the radial distribution function).
For the pure Lennard-Jones fluid under consideration here, we have
used $r_{\sigma}=0.9$ (in units of $\sigma$).

Once any possible initial overlap is sorted out ($U<U_{ovlp}$), the
second line of Eq. (\ref{ds}) is designed to drive the new particle
downhill in energy, towards the target value $U_0$.  Here resides the
main difference with respect to the reference scheme.  At large energies,
the typical slope of the potential energy is very large
($f^{(n)}>>1$), meaning that the energy drop along the steepest
descent direction is governed by the linear term of the Taylor
expansion in the displacement, $\Delta U= f^{(n)} \delta s +O(\delta
s^2)$.  The second line of Eq. (\ref{ds}) makes use of this fact and
takes $\Delta U=U^{(n)}-U_0$ for extracting a displacement adapted to
the (maximum) local energy gradient $\delta s= \Delta U/f^{(n)}$. Note
that at large energies $U^{(n)}-U_0 \sim U^{(n)}$, so after one
iteration one expects the energy to decrease in (at least) a fraction
of $U^{(n)}$, meaning a linear convergence.  The local curvature of
the potential energy landscape becomes dominant when approaching a
local minimum ($f^{(n)} \sim 0$) and in this case Eq. (\ref{ds}) limits
the displacement to a maximum value $\Delta s$. The maximum
displacement is the only variable parameter in the algorithm and, as discussed
in Sec. \ref{discusion}, its optimal value is about the width
of the low-energy tubes of the potential energy landscape (see Fig. 1)

At low energies,  as the iterator approaches the  energy target $U_0$,
the  displacement $\delta  s =  (U^{(n)}-U_0)/f^{(n)}$ behaves  like a
Newton-Raphson step made along the steepest descent direction.  Due to
this feature,  the convergence of the {\sc  usher} algorithm increases
notably  near the  target. In  particular, this  kind  of displacement
enables  the   error  $\xi$  to   decrease  quadratically  once
$\xi <  O(1)$. This fact is  illustrated in Fig.  2 by plotting
the  absolute value  of the  error $|\xi|^{(n+1)}$  against its
value  at the previous  iteration $|\xi|^{(n)}$.   As explained
above, for  $\xi >O(1)$  the algorithm converges  linearly with
$|\xi|^{(n+1)}\simeq     0.4     |\xi^{(n)}|$;     while
$|\xi|^{(n+1)}  \simeq  0.35 \left(  \xi^{(n)}\right)^2$
once  $\xi <O(1)$.   In the  same way,  it may  be  possible to
further increase  the convergence rate by  implementing a displacement
based upon  higher order  methods such as Halley's or  Bailey's scheme
\cite{hildebrand}  (for such purpose one  would need  to  calculate the
Hessian matrix and project it onto the steepest descent direction).

For the sake of completeness, we also describe here the older version of
the {\sc usher}'s displacement used in Ref. \cite{hybrid}.  This
earlier version used a similar displacement rule for $U^{(n)}>U_{ovlp}$
to that quoted in Eq. (\ref{ds}), but for lower energies it used
$\delta s^{(n)}= \min(\Delta s,\frac 12 f^{(n)} \Delta t)$ where the
optimal choice for the parameter $\Delta t$ ranged within $(0.05,
0.15)$. This scheme is around two times slower than the improved
{\sc usher} sheme discussed here.

One of the important issues of the algorithm design concerned the
optimal strategy to deploy for uphill iterations.  We compared two
different strategies. The first one, which we shall call {\em
indirect} {\sc usher} performs a line-minimisation \cite{numrec} of
the energy along the direction $\delta \mathbf{r}^{(n)}\equiv
\mathbf{r}^{(n+1)}-\mathbf{r}^{(n)}$, similar to that described in
Sec. \ref{ref} and Eq. (\ref{linmin}).  The second alternative, called 
{\em direct}-{\sc usher}, gives up the initial search and restarts a new
one from another random position $\mathbf{r}^{(0)}$ once an uphill
move is encountered. Interestingly the insertion tests (see
Sec. \ref{tests}) clearly show that the {\em direct} {\sc usher} is
about two times faster than the {\em indirect} version. This indicates
that most of the uphill moves encountered using the update rule of
Eqs. (\ref{ush}) and (\ref{ds}) are due to energy holes and therefore
suggests that Eq. (\ref{ds}) enables the {\sc usher} algorithm to
properly deflect its trajectory at most of the bends of the low
energy-tubes encountered.  A less restrictive version of the {\em
direct} {\sc usher} allows a line-minimisation iteration only if
the uphill move is done near enough to the target (for instance if
$|\xi|\leq O(1)$).  
This alternative gives slightly better results at large densities.

In the insertion tests presented below in Sec. \ref{tests} the
reference scheme is compared with the most efficient version of the
{\sc usher} algorithm; i.e. with the {\em direct} {\sc usher}.  To
avoid any possible confusion, in the remainder of the paper we shall
simply called this, the {\sc usher} algorithm.

\section{\label{tests} Insertion tests}

The insertion algorithms presented in Sec. \ref{des} were evaluated in
two kinds of systems, with and without periodic boundary conditions.
We stress that no thermostat was used in any of the
insertion tests.  This ensures that the temperature of the system does
not spuriously increase due to the dissipation of possible additional
internal energy introduced by particle insertions in non-appropriate
(higher-energy) locations.

In order to investigate the functioning of the insertion algorithm we
shall drive the system through a specific thermodynamic process
(see below) and compare the values of the thermodynamic variables
computed during the simulations with those arising from
thermodynamics.  The system contains $N$ particles within a volume $V$
and its total energy is $E=3NT/2+U(\mathbf{r}^N)$, the energy per
particle being $e=E/N$.  The thermodynamic processes will be specified
by the variation of the number of particles $\Delta N$ (or density
$\Delta \rho$) and the change of energy per particle $\Delta e$. We
now use standard thermodynamics to derive the changes in the system's 
other variables.

The variation of energy per particle upon insertion of $\Delta N$
particles into the system is $\Delta e =\Delta E/N - e \Delta N/N$.
For a system having no contact with a thermostat or a manostat, as for
the one considered here, the variation of the total energy upon
insertion of $\Delta N$ particles is exactly $\Delta E=
\left<\epsilon^{\prime}\right> \Delta N$, where
$\left<\epsilon^{\prime}\right> $ is the energy of the inserted
particle averaged over $\Delta N$ insertions; $\Delta N
\left<\epsilon^{\prime}\right>= \sum_i^{\Delta N} \epsilon^{\prime}_i$
( $\epsilon^{\prime}_i$ being the energy of the $i$th inserted
particle). Thus,
\begin{equation}
\label{delta_e}
\Delta e  =\left( \left<\epsilon^{\prime}\right>- e\right) \Delta N/N= \left( \left<\epsilon^{\prime}\right>- e\right) \Delta \rho/\rho.  
\end{equation}
The variation of temperature can now be calculated from the equation
\begin{equation}
\label{pro1}
\Delta T= \frac{1}{c_v} \left[\Delta e -\left(\frac{\partial{e}}{\partial \rho}\right)_T \Delta \rho\right] \\
\end{equation}
where $\Delta \rho =(\Delta N)/V$, $c_v$ is the specific heat at
constant volume and $(\partial{e}/\partial \rho)_T$ is obtained from
the equation of state for the excess energy per particle $u=U/N$
reported by Johnson {\em et al.} \cite{EOS}.

Therefore, for given initial values of the system's density and
temperature ($\rho_0, T_0$), the time evolution of the thermodynamic
variables is determined by the (specified) temporal variation of
density $\partial \rho/\partial t$.  The rate of temperature
variation, obtained from Eq. (\ref{pro1}), enables us to calculate the
temperature at each instant in the process.  The pressure and the
excess energy per particle can be then obtained from equations of
state ($P=P_{eos}(\rho,T)$ and $u=u_{eos}(\rho,T)$).

In the tests presented below we considered a thermodynamic process in
which the density increases at constant specific energy $\Delta e =0$. 
According to Eq. (\ref{delta_e}), during the process the average energy
of the inserted particles $\left<\epsilon^{\prime} \right>$ is set equal to the mean
specific energy of the system, $e$.  This condition is similar to that
required for the energy balance conditions in the hybrid
(particle-continuum) scheme of Delgado-Buscalioni and Coveney
\cite{hybrid}.  In fact, the process with $\Delta e=0$ can also be 
sought as a test for energy conservation in this hybrid scheme.
%We also considered processes with $\Delta T=0$. To that end particles were inserted 
%with energies $\left<\epsilon^{\prime} \right> =e+ \rho (\partial e/\partial \rho)_T$
%[see Eq. (\ref{pro1})].
%MODIFIED_NEW

The thermodynamic relations, such as Eq. (\ref{pro1}), are meaningful
at least under condition of local equilibrium.  This imposes a limit
on the rate of particle insertion, because within each subdomain of
the system the insertions of particles need to be sufficiently well
spaced out in time for the system to be able to recover the
equilibrium distribution.  Consider a small subvolume of size
$\lambda$,  large enough to be representative of the system's
distribution function.  For instance $\lambda$ may be the distance at
which the radial distribution function converges to one ($\sim
3\sigma$).  To be sure that the system is able to restore its
equilibrium distribution, the rate of particle insertion in each of
these subdomains $\lambda^3 (\partial \rho/\partial t)$ has to be
smaller than the inverse of the collision time, which for a simple fluid
can be estimated using the hard-sphere approximation by 
$\tau_c\simeq 0.14\rho^{-1} T^{-1/2}\,(\sigma^2 m/\epsilon)^{1/2}$
(see e.g., \cite{huang}). In our calculations we
used $(\partial \rho/\partial t)\le 0.01$, so the characteristic
insertion time was $\sim 3$, much larger than the collision time,
$\tau_c\simeq 0.3$

%Macroscopically, the rate of density increase
%$\rho^{-1}\partial \rho/\partial t$ should not exceed the inverse of
%the characteristic diffusive time of the system. An estimate of this
%time is $D/\lambda^2$, where $D$ is the mass diffusivity and $\lambda$
%the distance at which the radial distribution function tends to 1,
%$\lambda \sim 2.5\sigma$.  In the same way, local equilibrium impose
%certain conditions to the particle-continuum coupling.  The reader is
%refereed to Ref. \cite{hybrid} for a larger discussion of this issue.

\subsection{Insertions in a periodic box}
The first set of tests were performed in systems contained within a
cubic periodic box of side lengths $L=\left\{7,8,10\right\}\sigma$. The
initial density was set to a moderate value $\rho(t=0)=0.4$ and was
increased until $\rho\simeq 1.0$.  The maximum rate of density
increase used was $\partial \rho/\partial t \sim 0.01$.  The
temperature, pressure, excess energy per particle $u=U/N$ and total
energy per particle $e$ are plotted in Fig.3 {\em versus} the
density. Results correspond to particle insertions in a box with
$L=10\sigma$ at a constant density increase rate of $\partial
\rho/\partial t=0.01$.  Particles were inserted at sites where the
potential energy equals the specific excess energy of the system
$U_0=U/N=u$ and with velocities drawn from a Maxwellian distribution
at the instantaneous system's (kinetic) temperature $T$.  The dashed
lines in Fig. 3 correspond to the thermodynamic variables obtained
from the the equation of state according to the process of
Eq. (\ref{pro1}).

\subsection{Insertions in an open flow}

The second test was done in open fluid flows, i.e., in systems with
open boundary conditions. We considered a system with density $N/V$
within a cuboidal domain of sides $L_x=40\,\sigma$ and $L_y=L_z=9\, \sigma$. The system
is periodic in the $y$ and $z$ directions and has open boundaries at
$x=0$ and $x=L_x$.  Particles were inserted with potential energies
close the specific excess energy $U_{eos}/N$ and with velocities drawn
from a Maxwellian distribution at a temperature $T_0$ which was fixed
throughout the simulation.  Insertions were done within a region of
width $\Delta x$ around $x=0$ and at a rate $A\rho v_{in}$, 
where $v_{in}$ is a parameter that determines the flow
velocity normal to the surface vector of the open boundary
($\mathbf{n}$) and $A=L_y L_z$ is the area of the boundary. At the
right hand boundary, particles are extracted at the same rate, so the
overall density of the system remains constant throughout the
simulation $N/V$.  In order to couple the particle region to the outer
pressure we used our hybrid particle-continuum scheme at the $x=0$ and
$x=L_x$ surfaces \cite{hybrid}.

Note that in this case particle insertion in the
$x$-direction is restricted to a region $\Delta x$ which is set to
$\Delta x=2.0\sigma$ (the volume available to insert particles being
$\Delta x L_y L_z$).  To ensure that insertions are done in this
region two different strategies were implemented. The first one is a
simple reflection of the position $\mathbf{r}^{(n+1)}$ back to the
insertion domain when the {\sc usher} iterator crosses the $x=0$ and
$x=\Delta x$ boundaries.  In the second implementation we imposed an
artificial sharp potential well at $x=0$ and $x=\Delta x$ which acts
only during the evaluation of the forces in the iteration procedure,
i.e., it was not included in the evaluation of
$U(\mathbf{r}^{(n+1)})$.  Both alternatives worked equally well and
resulted in a similar number of required iterations.

The simulation starts from an initial state with zero mean velocity
and constant density profile along the $x$-direction.  As time goes
by, the particle insertions concentrated in the region around $x=0$
lead to the production of a density wave that expands at the sound
velocity for $x>0$.  This density wave transports momentum along the
$x$-direction and, after a transient time, the density profile
converges to the flat stationary density profile; throughout the
simulation cell, the mean flow $x$-velocity tends to the value $v_{in}$.

The hydrodynamic and thermodynamic variables were measured over slices
of width $\Delta x$ along the $x$ direction. Figure 4 shows the local
density at some of the left most slices $x < L_x/2$ together with the
mean (slice averaged) velocity and the total temperature of the
system. The oscillatory behaviour of the local density is a desired
feature of these tests as it enables us to determine the dependence of
the number of iterations $n$ on the density for a range of values of
$\rho$ in each simulation.  We refer to our previous
paper\cite{hybrid} for a detailed comparison between theoretical
hydrodynamic trends and results obtained from hybrid
continuum-particle simulations in different relaxing flows, also
involving mass exchange.

\section{\label{results}
Results}

Figure 5 presents the average number of single-force evaluations
$\left<n_{f}\right>$ needed to perform an insertion versus the
density.  The most expensive part of the insertion algorithms is the
evaluation of the force.  Consequently in order to compare the
performance of the algorithms we have used $n_f$, rather than the
total number of iterations $n$.  We note that some of the steps
discussed in Sec. \ref{des} (such as line-minimisation) require
several sub-iterations and so $n \leq n_f$.  The {\sc usher} and the
reference algorithms are compared in Fig. 5, in a test corresponding to
insertions in a periodic box.  In this kind of test (and for a similar
average error $\left<\xi\right> <0.05$) the {\sc usher}
algorithm is more than two times faster than the reference scheme for
$\rho>0.5$ and more than four times faster for $\rho>0.8$. The
reference scheme is slightly slower when insertions are 
constrained to a smaller region, as occurs in the open fluid flow
tests. But notably, for both open flow and periodic box tests, the
{\sc usher} scheme gives similar values of $\left<n_{f}\right>$ . This
means that the {\sc usher} algorithm does not pay any extra cost for
restricting the size of the domain of insertion.  This may be
understood by looking at the distance between the initial trial
and final insertion positions, $\Delta r =
|\mathbf{r}^{(0)}-\mathbf{r}^{(n)}|$, shown in Fig. 6.  For a wide
range of densities the maximum value of $\Delta r$ is 
smaller than 1.0$\sigma$ (its average being typically less than
0.5$\sigma$), indicating that most particles are inserted before the
{\sc usher} iterator reaches the boundaries of the insertion
domain. This feature important for many
applications. For instance, in the hybrid particle-continuum schemes
the insertions are assigned (and restricted) to finite cells arising
from a discretisation of the space \cite{Flek00,hybrid}; and in the
water-insertion method \cite{waterins}, the water molecules have to be
placed in assigned protein cavities.

Figure 5 illustrates how the number of force evaluations varies with
the maximum averaged error when using the {\sc usher} algorithm.  In
particular we compare the results for the insertions tests done at an
initial temperature of $T_0=3$ (as in Fig. 2) and with different
values of the maximum error averaged over 30 insertions, $\left<\xi
\right>$.  To decrease the error, from $0.15$ to $0.03$ one typically
needs one more iteration. Another iteration leads to
$\left<\xi\right>=10^{-3}$. This fast (quadratic) error
reduction is made possible by the Newton-Raphson-like displacement
implemented in Eq. (\ref{ds}) (see Fig. 2).  A systematic
non-vanishing value of $\left<\xi\right>$ has a direct effect
on the thermodynamic variables, as shown in Fig. 3.  For instance, a
value of $\left<\xi\right>\simeq 0.05$ maintained during the
insertion process leads to a systematic drift from the $\Delta e=0$
line, and also has an effect on the temperature evolution.

%MODIFICATION
Additionally, Fig. 5 illustrates how $\left<n_f\right>$ varies with
the system's temperature as shown in the data for $T_0=3$ and
$T_0=10$. At larger temperatures it becomes much easier to insert
particles once $\rho>0.6$. The reason is that the target energy
$U_0(=u_{eos}(\rho,T)$) increases much faster with the temperature at
larger densities than it does at lower densities.  For instance,
$u_{eos}(0.4,3)\simeq -1.9$ and $u_{eos}(0.4,10)\simeq -1.2$, while
for a larger density $u_{eos}(0.85,3)\simeq -3.1$ and
$u_{eos}(0.85,10)\simeq -0.6$.  For the same reason, if particles were
inserted with potential energies similar to the chemical potential
$\left<U_0\right>=\mu_{eos}(\rho,T)$, the slope of $\left<n_f\right>$
with $\rho$ would be flatter than those data shown in Fig. 5. We also
performed insertions at subcritical temperatures $T<1.3$, for liquid
densities and also inside the liquid-vapour coexistence region. In these calculations
the number of iterations needed to insert a particle was very
similar to that presented in Fig. 5 for $T_0=3$.  Anyhow, fluctuations
of $n_f$ were larger inside the coexistence region as a consecuence of
the inhomogeneity of the density field.

%Processes with $\Delta T=0$ (``isothermal'' insertions ) were performed by
%inserting particles with $\left<\epsilon^{\prime}\right>=e+ (\partial
%e/\partial \rho)_T$ [see Eq. (\ref{pro1})]. A calculation at
%subcritical temperature, T=0.9, is shown in Fig. 5.  The temperature
%fluctuates arround its initial value and the pressure remains constant
%while the fluid is in the liquid-vapour coexistence region, $\rho \in
%(0.05,0.63)$. Again, we state that no thermostating was used in any of
%the insertion tests presented here.
%MODIFIED_NEW

\section{\label{discusion}
Choosing an optimal parameter set}

We now wish to provide a physical interpretation of the performance of
the particle insertion algorithms, based on the structure of the
energy landscape.  Such insight will be very useful for extensions of
the {\sc usher} parameters to the simulation of other kinds of
fluids. In fact, our experience is that, instead of a simple
parametric study, it is advisable perform an analysis of the structure
of the potential energy landscape to obtain information about the
typical shapes and length scales of the low energy regions.  This kind
of structural analysis for the Lennard-Jones fluid considered here not
only provided important clues for the algorithm design, but provided
the key relationship between the optimal displacement $\Delta s$ and
the density.

\subsection{Low energy holes}
In order to investigate the structure of the low energy holes we
devised the following procedure.  In a standard MD simulation in a
periodic box and at time intervals separated by several collision
times, we seek a point $\mathbf{r}_0$ with a very low pre-specified
energy $U_0$.  In particular $U_0$ is chosen to be the mean excess
energy per particle.  Initially the search for the point
$\mathbf{r}_0$ was done by the ``basic'' update rule of the {\sc
usher} algorithm mentioned in Sec. \ref{usher}.  Once $\mathbf{r}_0$
was found, the energy landscape was probed in radial directions from
this point. For each azimuthal angle $\psi\in \left[0,2\pi\right]$ and
longitudinal angle $\theta \in \left[0,\pi \right]$, the energy
$U(\mathbf{r})$ was measured for increasing radial coordinate and a
radial distance was recorded when $U(\mathbf{r})\geq U_{iso}$.  The
radial distance will be denoted as $R(\psi,\theta)$.  Note that
$R(\psi,\theta)$ determines the shape of each hole; in particular the
mean radius and the mean of the squared radius were computed for each
hole:
\begin{eqnarray}
\label{meanR}
\left<R\right>_{\theta,\phi}\equiv\frac{1}{2\pi^2}\int_{0}^{2\pi} d\phi  \int_{0}^{\pi}  R(\theta,\phi)d\theta, \\
\label{meanR2}
\left<R^2\right>_{\theta,\phi}\equiv\frac{1}{2\pi^2²}\int_{0}^{2\pi} d\phi  \int_{0}^{\pi}  R^2(\theta,\phi)d\theta.
\end{eqnarray}
The effective shape of the hole can be estimated by the following quantity,
\begin{equation}
\label{varR}
\sigma_R\equiv \left(\frac{\left<R^2\right>_{\theta,\phi}}{\left<R\right>_{\theta,\phi}^2}-1\right)^{1/2}.
\end{equation}
Clearly for a sphere $\sigma_R=0$, while $\sigma_R$ is positive for
any other elongated shape.  A glance at the low and intermediate
energy regions of a typical contour plot of the potential energy (see
Fig. 1) suggests that it is possible to estimate the characteristic
length scales of the low energy regions by fitting $\sigma_R$ and
$\left<R\right>_{\theta,\phi}$ to ellipsoids.  In particular, due to
the symmetry of the LJ fluid, it is enough to use asymmetric
ellipsoids for this estimation. For an ellipsoid with semi-minor and
semi-major axes given respectively by $R_s$ and $R_l=e\,R_s$, the
following parametric relations fit within $1\%$ to the exact
analytical results:
\begin{eqnarray}
\label{sigel}
 \sigma_{R}=0.56\log \chi,\\
\label{Rel}
\left<R\right>_{\theta,\phi}=R_s(1+0.25\log \chi).
\end{eqnarray}
For given values of $\sigma_R$ and $\left<R\right>_{\theta,\phi}$, one
can estimate the eccentricity $\chi=R_l/R_s$ and the semi-minor axis
$R_s$ using Eqs. (\ref{sigel}) and (\ref{Rel}) .  The values of
$\left<R\right>_{\theta,\phi}$, $\sigma_R$ and the estimations of
$R_s$ and $R_l$, averaged over a set of about 80 holes, are shown in
Fig. 7 {\em versus} the density.  To ensure that these values are
representative of the shape of low and intermediate-energy regions
(such as those shown in Fig. 1), the averages were obtained for a
relatively wide range of (intermediate) energy isovalues $U_{iso} \in
[0,50]$. The error bars determine the maximum variation of these
quantities for this range of $U_{iso}$.

In Fig. 7 we included the optimum choice for the reference algorithm
described in Sec. \ref{ref} ($\Delta s_1$ and $\Delta s_2$) .  The
optimisation of these parameters was performed independently, for a
wide range of densities, $\rho \in [0.4-0.92]$ (see Fig. 8). It is
interesting to note that $\Delta s_1$ closely follows the trend
obtained for the smallest effective radius $R_s$, while $\Delta s_2$
lies above the longest radius $R_l$ of the intermediate energy
regions.  The interpretation of the results for $\Delta s_1$ seems
quite evident, meaning that the displacement of the steepest descent
method when moving downhill should be about half the minimum
characteristic diameter of the low energy valleys.  From this we
readily understand why the best choice for the maximum displacement of
the {\sc usher} algorithm is $\Delta s \simeq R_s\simeq
0.1\rho^{-1.5}$. Quite remarkably, the estimate of the mean free path
based on the hard-sphere fluid, $0.2 \rho^{-1}$ (shown in Fig. 7), is
close to the typical radius of the low energy holes
$\left<R\right>_{\theta,\phi}$.  This indicates that such kinetic
information, if available, may be of great help for the first
adjustment of the maximum displacement of the algorithm, when
inserting particles (or minimising the energy) in other kinds of
fluids.

Fig. 8b sheds light on the interpretation of the result for $\Delta
s_2$.  The optimal value of $\Delta s_2$ may be taken to be any value
larger than a certain threshold, which according to Fig. 8b has to
exceed the largest typical longest diameter within the low energy
regions. This confirms that once an uphill move is made the fastest
option is to completely traverse the energy valley and continue the
iterations from a high energy site, instead of trying to pursue
possible further line minimisations.  This conclusion, obtained from
the reference scheme, suggested that the best procedure was to give-up
the search once $U^{(n)}>U^{(n-1)}$. As stated in Sec.  \ref{usher},
this is indeed what we have found when comparing the {\em direct} and
{\em indirect} versions of the {\sc usher} scheme.

\section{\label{con} Conclusions}

An increasing number of methods involving molecular dynamics (MD)
simulations of open systems \cite{Pet92,Pet_more,waterins,Flek00}, require one
to insert particles at precise locations where the potential energy is
set equal to a pre-specified value. Moreover, insertions need to be
done on the fly and the performance of these methods will greatly
depend on the efficiency of the insertion algorithm.  At high
densities this may seem a formidable task and indeed this sort of
insertion algorithms has scarcely been explored in the literature.
The main purpose of our paper is to show that this problem can be
efficiently accomplished. To this end we have devised a particle
insertion procedure called the {\sc usher} algorithm.  To give an
example, to insert a particle in a Lennard-Jones fluid with $\rho =
0.5$ and $T=3.0$, at positions where the potential energy equals the
mean specific energy of the system, the algorithm requires around 8
extra evaluations of a (single-particle) force and 25 if $\rho = 0.8$.

The {\sc usher} algorithm essentially consists of a steepest descent
iteration procedure (see Eq. \ref{ush}) with a displacement adapted to
the local shape of the energy landscape.  In particular, by using an
initial displacement which depends on the value of the potential
energy at the initial trial position, $1-(4/U^{(0)})^{1/12}$, the
algorithm avoids in (about) one iteration any possible overlap with a
pre-existing particle.  We confirmed that this feature makes it
advantageous to choose the initial trial position at random, instead
of using a much more expensive grid method to slice up the entire
space in the search of the less dense region (as is done in the
cavity-biased Monte-Carlo or in the Grand-Canonical MD scheme proposed
by Pettitt and coworkers \cite{Pet92,Pet_more}). In subsequent iterations the
displacement is given by the Newton-Raphson step measured along the
steepest descent direction and has a upper bound of $\Delta s$, 
to avoid uncontrolled jumps near local minima.

As another relevant conclusion, we wish to caution about the usage of
line minimisation, normally implemented in conjunction with the
steepest descent method \cite{numrec,MDbook}.  We clearly observed
that, in these complex landscapes, it is better to use a (small
enough) maximum displacement to ensure that most the iterations are
made downhill; then, if a single iteration is made uphill, the best
option is to restart the search from another random position, rather
than performing a line minimisation.  There are two reasons for this
fact: firstly, line minimisation is expensive and secondly, and more
importantly, we observed that if the maximum displacement is optimal,
most of the uphill moves are due to the presence of an energy trap
(i.e., a local minimum at an energy larger than the target).

An important part of this work was to give a physical interpretation
of the optimal maximum displacement $\Delta s$.  As stated earlier it
is essential to optimise its value for the complex topology of the low
energy-tubes.  To that end we analysed the structure of the energy
landscape of the Lennard-Jones fluid considered here and concluded
that the width of low-energy tubes scales as $0.1\rho^{-1.5}$.  A much
more computationally expensive parametric study clearly showed that
the optimal displacement follows a similar trend. This means that, in
order to extend the insertion algorithm to other fluids, it is
strongly advisable to first investigate the structure (in terms of
shape and length scales) of the low energy regions.  Apart from this,
our insertion protocol is based uniquely on mechanical variables
readily calculated in any standard MD simulation (force and potential
energy), therefore the same kind of protocol can be used for inserting
solvent molecules in fluids consisting of (small) poly-atomic
molecules or even in polar fluids with nonadditive potentials.
Nevertheless the optimal algorithm design may require modifications,
depending on the specific molecule considered.  For instance, in
fluids whose molecules have rotational degrees of freedom, the
insertion update step could be modified so as to first update the position
of the center of mass of the molecule and then to use the local torque
to orientate the molecule to the most favourable position. Such an
investigation will form the subject of future work.
%MODIFICATION

\section{Acknowledgements}

We gratefully acknowledge useful discussions with E. Flekkoy and
P. Espa\~nol. RD-B wishes to thank G. Ciccotti and R. Winkler for 
fruitful discussion and S. Saumitra for comments. This work is
supported by the European Union through a Marie Curie Fellowship
(HPMF-CT-2001-01210) to RD-B. Support from the project BFM2002-0290 is
also acknowledged.

\newpage 

\section{Figures}
\begin{itemize}
\item Figure 1\\
 (a) A cut along the x=0 plane of the contour plot of the
potential energy landscape for energies lower than $100$, showing the
typical low-energy tube-like structures.  In (b) a close up of the
left most region indicating with thicker solid lines a possible
targeted energy, at $U = -4$.  Some bends and saddle points of the
energy surface and some energy traps (local minima with energy larger
than the target) are indicated with solid and dashed arrows,
respectively.  The snapshot corresponds to a LJ fluid with $\rho=0.6$
and $T=2.5$ inside a 3D cubic box of side $L=10\,\sigma$.  The size of the
maximum displacement $\Delta s$ used by the {\sc usher} algorithm for
this density is also indicated.

\item Figure  2\\
The absolute value of the error at the $n+1$ plotted against
its value at the previous iteration $n$. The data corresponds
to insertions made by the {\sc usher} algoritm
in a periodic box $L=10\,\sigma$ and for
$0.6\leq \rho\leq 0.75$. As illustrated by the dashed lines,
the convergence is linear for $\xi>O(1)$
and quadratic for $\xi<O(1)$.

\item Figure 3\\
 (a) Total energy per particle $e$, (b) pressure $P$ , 
(c) temperature and, (d) excess energy per particle $u$ 
{\em versus} the density, obtained in a particle insertion test 
made in a cubic periodic box with side length $L=10\,\sigma$. 
The density increases linearly
with time at a rate of $\partial \rho/\partial t=0.01$.
The insertions were made to guarantee a process with $\Delta e=0$
(see text). The dashed lines corresponds to the thermodynamic 
variables extracted from Eq. (\ref{pro1}),
using the equations of state for $u_{eos}(\rho,T)$ 
and $P_{eos}(\rho,T)$.

\item Figure 4\\
Evolution in time of various hydrodynamic
variables in an insertion
test on an open flow in a ``box'' with sides $L_x=40\,\sigma$, $L_y=L_z=9\,\sigma$.
Particles were inserted with zero mean velocity at $x=0$ at a rate
$s=L_zL_y\rho_0 v_{in}$ and extracted at $x=L_x$ at the same rate.
The overall density was $\rho_0=0.5$ and $v_{in}=1.0$.  (a) Mean
and local
temperature and (b) local density at slices of width $\Delta x=2$; in
(c) the local velocity at each slice (dotted line) and the mean
velocity (solid line) are shown. The flat  stationary $x$-profiles 
of density ($\rho=0.5$) and velocity ($v_x=v_{in}=1.0$)
are reached after several sound transversal periods.

\item Figure 5\\ 
The average number of force evaluations needed to insert a
new particle $\left<n_f\right>$. The results correspond to insertions
in a cubic periodic box of side $L=10\,\sigma$.  For all the curves
$\left<\xi\right>$ indicates the maximum value of the averaged errors
and the error bars corresponds to the standard deviation upon 100
insertions.  The results are for processes with $\Delta e=0$, starting
from an initial temperature $T_0$ [see Eq. (\ref{pro1}) and Fig. 3].

\item Figure 6\\
The distance travelled by the {\sc usher} algorithm between
the initial trial position and the final insertion site, $\Delta
r\equiv |\mathbf{r}^{(0)}-\mathbf{r}^{(n)}|$, (in log-scale) {\em
versus} the density. The test corresponds to particle
insertions in a cubic periodic box of side $L=10\,\sigma$.

\item Figure 7\\
 The mean radius of the low energy regions
$\left<R\right>_{\theta,\phi}$ [Eq. (\ref{meanR})] along with the
smallest $R_s$ and largest $R_l$ characteristic lengths of the low
energy regions estimated by Eqs. (\ref{Rel}) and (\ref{sigel}).
Squares correspond to the optimum values of the reference 
scheme $\Delta s_1$ obtained from a parametric study. 
The dashed line ($0.1 \rho^{-1.5}$) 
corresponds to our choice for the optimum {\sc usher} maximum displacement
$\Delta s$. The mean free path
(hard-spheres estimate $0.2\rho^{-1}$) is also shown. The inset
shows the normalised variance $\sigma_R$ given by Eq. (\ref{varR})
and the estimated mean eccentricity
of the low-energy holes $\chi=R_l/R_s$.

\item Figure 8\\
 The average number of force evaluations per insertion {\em versus}
the parameters (a) $\Delta s_1$ and (b) $\Delta s_2$ of the
reference scheme of Sec. \ref{ref}. The evaluations were done by
inserting particles in a cubic periodic box of side $L=10\,\sigma$.
The range of densities at which the evaluations
were made is indicated in each figure.
\end{itemize}

\end{document}